\documentclass{article}
\usepackage{spconf,amsmath,graphicx}
\usepackage{amssymb}

\title{Iterative Gradient Encoding Network with Feature Co-Occurrence Loss for Single Image Reflection Removal}
\name{Sutanu Bera, Prabir Kumar Biswas }
\address{Department of Electronics and Electrical Communication Engineering, \\Indian Institute of Technology Kharagpur}
\begin{document}
%\ninept
%
\maketitle
\begin{abstract}
Removing undesired reflections from a photo taken in front of a glass is of great importance for enhancing visual computing systems' efficiency. Previous learning-based approaches have produced visually plausible results for some reflections type, however, failed to generalize against other reflection types. There is a dearth of literature for efficient methods concerning single image reflection removal, which can generalize well in large-scale reflection type. In this study, we proposed an iterative gradient encoding network for single image reflection removal. Next, to further supervise the network in learning the correlation between the transmission layer features, we proposed a feature co-occurrence loss. Extensive experiments on the public benchmark dataset of SIR$^2$  demonstrated  that our method can remove reflection favorably against the existing state of the art method on all imaging settings, including diverse backgrounds.  Moreover, as the reflection strength increases, our method can still remove reflection even where other state of the art methods failed. 

\end{abstract}
\section{Introduction}
\label{sec:intro}
How to obtain a reflection-free image taken through the glass is of great interest to computer vision researchers. Removing the undesired reflection enhances the target object's visibility and benefits various computer vision tasks, such as image classification, segmentation, and object detection. The initial efforts of reflection removal employed multiple images to disentangle the reflection from the transmission \cite{guo2014robust}, \cite{sarel2004separating}, \cite{szeliski2000layer}. More recently, the endeavor to solve the more common and practically significant scenario of a single input image has received a lot of appreciation \cite{li2014single}, \cite{shih2015reflection}, \cite{li2020single}. However, single-image reflection removal is a challenging process because of the ill-posed nature of the problem \cite{levin2007user}. To deal with the problem's ill-posedness, the recent learning-based method has utilized different auxiliary information as prior and constraint \cite{fan2017generic}, \cite{wan2018crrn}, \cite{lyu2019reflection}, \cite{sun2019multi}. Among them, one school of researchers had tried to exploit auxiliary information embedded in the image's gradient \cite{abiko2019single}, \cite{wan2019corrn}. For example, in \cite{wan2018crrn} Wan et al. proposed to use an auxiliary network to restore the gradient of the corrupted image and fused the features of the gradient network with the image restoration network. In a different work, M Ikehara et al.\cite{abiko2019single} proposed a feature space gradient constraint loss. In \cite{wan2019corrn} Ket et al. proposed an MMD loss based on the higher-order statics of image gradient. Together with other learning-based methods, these methods perform adequately when the effect of reflection is low; however, they fail miserably if the intensity of reflection increases. Moreover, due to different complexities such as insufficient training data, varying imaging conditions, diverging scene content, and little physical understanding of this problem, these methods often fail to remove some reflection but perform sufficiently for another reflection type.
In this study, we proposed a novel iterative gradient encoding network for single image reflection removal. Our proposed iterative method first computes the gradient of the estimated transmission layer of the current iteration and utilizes it to estimate the transmission layer in the next iteration. We have employed a multi-scale feature fusion scheme to use the feature of gradient image. Note, our proposed method is different from other iterative methods, as we have not used the estimated transmission as an input to the network in the next iteration; only a subpart of the network is iterated, holding the feature extracted from the mixture image fixed.
\\Next, we reintroduced the well-known style loss \cite{gatys2015neural} as feature co-occurrence loss for reflection removal. The correlation between the features of the transmission layer and the reflection layer provides useful information for separating these layers. Style loss is well known for capturing correlation between different features, yet this consequential but straightforward loss is unused in reflection removal. To the best of our knowledge, this is the first study to adopt style loss as feature co-occurrence loss for reflection removal. 
\\We evaluated our proposed method in the recent benchmark data set of SIR$^2$ \cite{wan2017benchmarking}. This data set contains real-world reflection images with diverse imaging settings and backgrounds. Extensive experiment on this data set has shown our method can remove reflection more efficiently than the current state of the art in all imaging conditions. Moreover, our iterative methods have helped to remove strong reflections from the background where most of the state of the art method failed.
\begin{figure*}[h]
    \centering
    \includegraphics[width=0.8\textwidth]{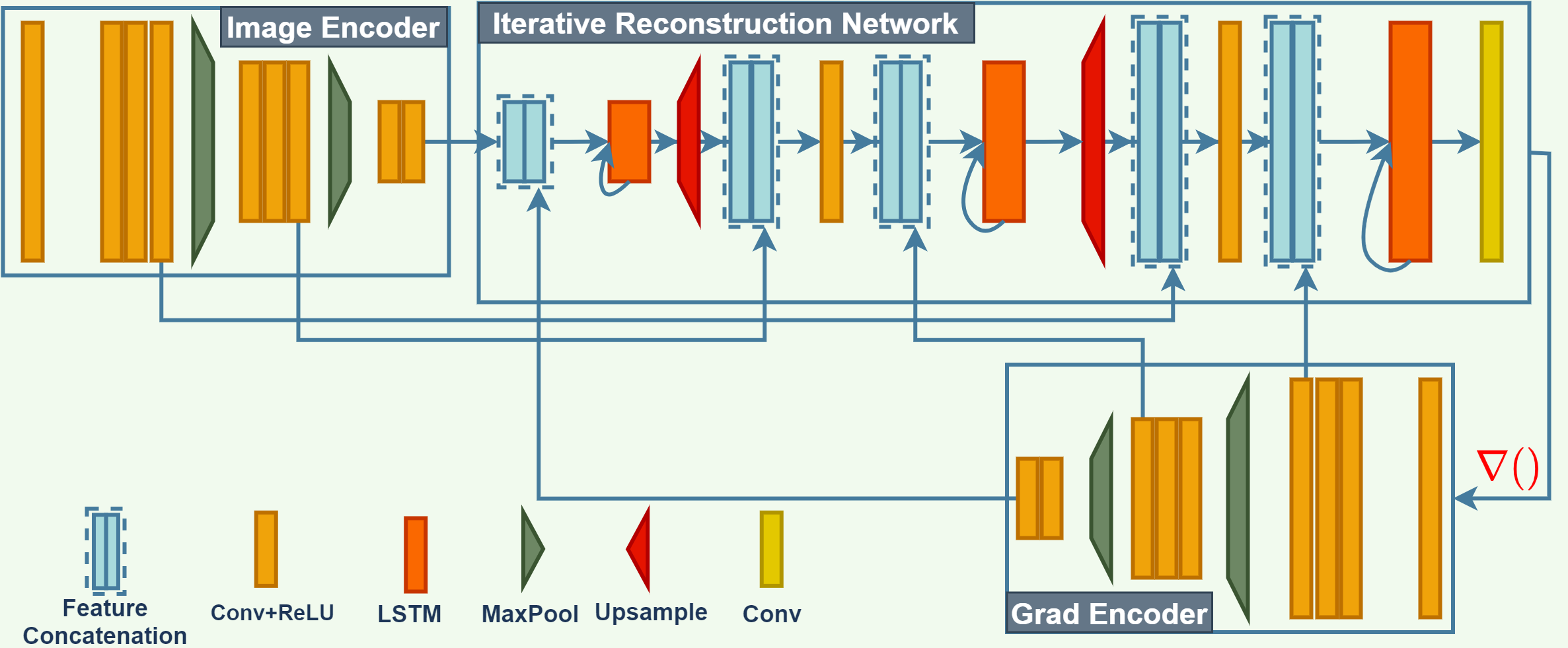}
    \caption{Our Proposed Iterative Gradient Encoding Network}
    \label{fig:igen}
\end{figure*}
\section{Proposed Method}
\label{sec:method}

Gradient provides vitally important clues for separation of reflection from the background. In general, the gradient of the transmission layer contains larger values, and the reflection layer is with smaller gradient values. So the pixel belonging to the transmission layer can be more easily differentiated in the gradient domain. In this work, we proposed an iterative gradient encoding network to encode the estimated transmission layer's gradient image and use these features to identify the pixel location belonging to the transmission layer in the next iteration. Our proposed iterative gradient encoding network is shown in Figure \ref{fig:igen}. Our proposed network comprise of three main components: image encoding sub-network $\phi$, gradient encoding sub-network $\theta$, and iterative image reconstruction network $\psi$.
The iterative reconstruction network's output is the estimated transmission layer $\hat{T}$. At an iteration step t, the output $\hat{T}(t)$ is given by,
\begin{center}
    $\hat{T}(t)=\psi\Big(\phi(I);\theta\big(\nabla \cdot \hat{T}(t-1)\big)\Big) $
\end{center}
Where $I$ is the original mixture image and $\nabla$ is the gradient operator. We initialize the $\hat{T}(0)$ as $I$. Note, the image encoding network's output and input are fixed across all iterations; at every iteration, these encoded features are being mixed with the updated gradient features. We speculated that the gradient would be more prominent at every iteration. The gradient features evolution will guide the iterative reconstruction network to identify the pixel belonging to the transmission and interpolate the remaining pixel based on these pixels. To effectively learn the evolution of gradient features as well as the long-range dependency among gradient features and image features, we have utilized the well-known convolutional LSTM unit. 
\\ The efficacy of multi-scale representation for reflection removal has already been well studied in the previous literature \cite{wan2018crrn}, \cite{wan2016depth}. Inspired by these works, we have used a novel multi-scale feature fusion scheme to fuse gradient image features with the original image. As shown in Figure \ref{fig:igen}, we have used three different scales for feature fusion; at every scale, first, the features of the image encoder are mixed with the interpolated feature by a convolution layer, these features are concatenated with the updated gradient feature and then forwarded to a two-layer convolutional cell for interpolation followed by upsampling. In every convolutional layer and convolutional LSTM cell, we have used $3 \times 3$ kernel.\\
\textbf{Feature Co-Occurrence Loss:}
In the mixture image, objects from transmission layers and reflection layers are overlaid with each other. However, in the original transmission layer, this type of superposition does not happen. In this study we proposed to use feature co-occurrence loss to regulate this type of superimposition. It is defined as:
\begin{center}
    $\mathcal{L}_{f}=\sum_l ||G\big(\zeta^l(\hat T \big)-G\big(\zeta^l(T\big)||_2^2$
\end{center}
with Gram matrix $G(F)=FF^T \in \mathbb{R}^{ n \times n}$, and $\zeta^l(I) \in \mathbb{R}^{n \times m}$ is the feature activation at the $l^{th}$ pre-trained VGG layer \cite{simonyan2014very} with n features of length m. Note, our objective is not to transfer the original transmission image's style content but to restrict unrealistic feature co-occurrence in the estimated transmission image by forcing the correlation among the feature extracted from the estimated transmission image to resemble the feature correlation extracted from the original transmission image.\\
\textbf{Other Loss Function:}
We have also used Mean Absolute Error/ L1 Loss ($\mathcal{L}_{1}$), perceptual loss ($\mathcal{L}_{p}$), and adversarial loss  ($\mathcal{L}_{adv}$) between real transmission layer ($T$) and estimated transmission layer ($\hat{T}$)  to train our network. So, our total loss function $\mathcal{L}_{t}$ is given as:
\begin{center}
     $\mathcal{L}_{t}= \mathcal{L}_{1}(\hat{T},T) + \lambda_1 \mathcal{L}_{p}(\hat{T},T)+ \lambda_2 \mathcal{L}_{f}(\hat{T},T)+ \lambda_3 \mathcal{L}_{adv}(\hat{T},T)$
\end{center}
We empirically set the values of weight $\lambda_1$, $\lambda_2$, $\lambda_3$ as 0.1, 0.1, 0.5. Details about these loss functions and architecture details of discriminator network is given in the supplementary material.
\begin{figure*}[h]
    \centering
    \includegraphics[width=\textwidth]{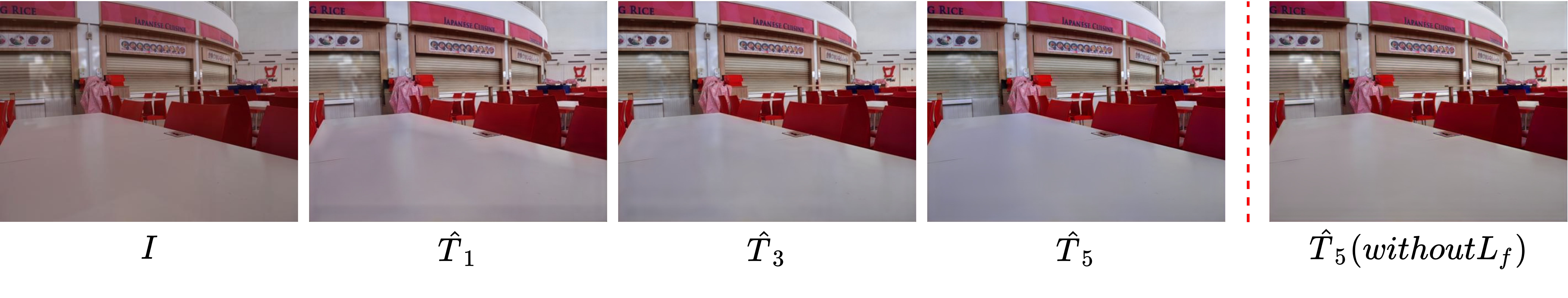}
    \caption{Visualization of results at different iteration evaluated on the Wild Screen Dataset of SIR$^2$.  The estimates of transmissions become increasingly more accurate with increase in the iteration. More results are on the supplementary material.}
    \label{fig:abla}
\end{figure*}
\section{Experimental Details:}
 We have used the recently proposed Reflection Image Dataset (RID) \cite{wan2019corrn} as the training set. Also followed the similar procedure as given in \cite{wan2019corrn} to synthesize the reflection images. For training our network, we used the whole images instead of the patch-based training strategy. For optimization, we used Adam optimizer with a batch size of 16. The remaining training details are given in the supplementary material. For computing gradient, we have used the following three filter on each of the RGB channels, $f_{1}= [-1\ 1], f_{2}=[-1\ 1]^{T}, f_{3}=\begin{bmatrix} 
0 & 1 & 0 \\
1 & -4 & 1\\
0 & 1 & 0\\
\end{bmatrix}$.
\\For evaluation of our proposed method, we have used the recently proposed benchmark data set SIR$^2$ \cite{wan2017benchmarking}. This data set contains real-world reflection images with diverse backgrounds and imaging conditions. For comparison we considered following state of the art methods, PNet\cite{zhang2018single},  ERRNet \cite{wei2019single}, GCNet\cite{abiko2019single}, IBCLN\cite{li2020single}, CoRRN\cite{wan2019corrn}. For a fair comparison, we have used the codes and trained model provided by the original authors.
\label{sec:pagestyle}

\section{Result and Discussion}
\label{sec:result}
\begin{figure*}[h]
    \centering
    \includegraphics[width=\textwidth]{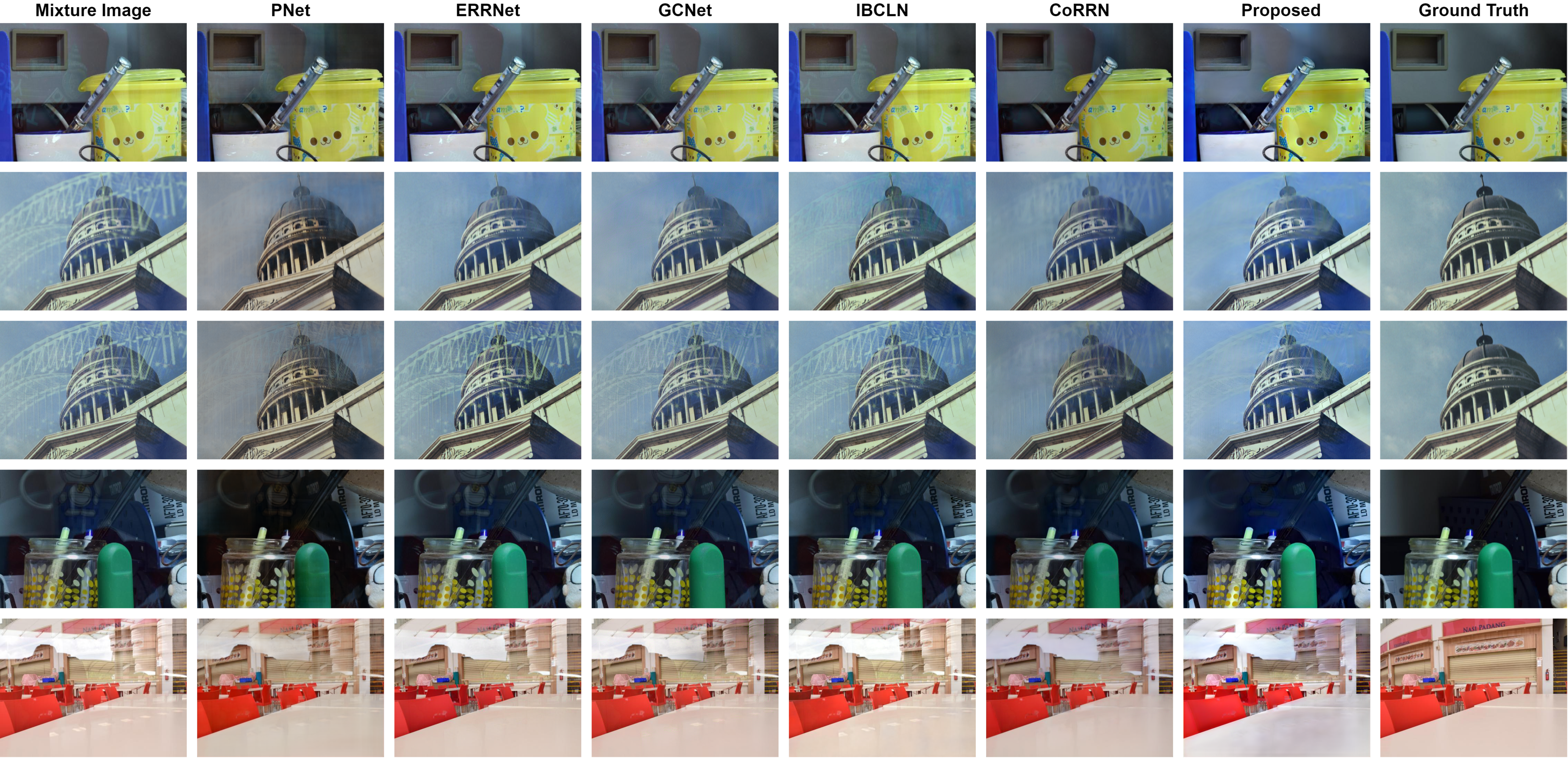}
    \caption{Examples of reflection removal results evaluated on the SIR$^2$ Dataset. Viewers are encouraged to zoom in for better view. More results are on the supplementary material.}
    \label{fig:sota}
\end{figure*}
In this section, we first show the efficacy of the proposed iterative reconstruction network. In Figure \ref{fig:abla} we have given an example reflection image taken from wild screen subset of SIR$^2$ dataset, and estimated transmission image at different iteration is shown on the right-hand side. The effect of the reflection is becoming more debilitated as the number of iteration is increasing. In the rightmost image of Figure \ref{fig:abla} we have given the result of reflection removal of the same network but trained without feature co-occurrence loss. The reflection in the shutter of the left side shop is still present. The feature co-occurrence loss forces the network to regulate this type of concurrent feature appearance; thus, reflection removal is more accurate. Moreover, the feature space loss also provides a global consistency in the generated image; thus, the table has obtained its original white color. Whereas in the output of the model without feature co-occurrence loss has a reddish appearance.  We have added more results in the supplementary material with different imaging conditions and backgrounds to observe the benefit of the iterative reconstruction network and feature co-occurrence loss. Next, following the de facto practice in the literature, we have compared our reflection removal result with other state of the art methods objectively in Table \ref{tab:table} in terms of PSNR and SSIM. Our method has achieved the best PSNR and SSIM among the current state of the art methods like PNet, ERRNet, and GCNet, CoRRN. Next, we performed a blind reader study for subjective evaluation. We asked five volunteers to rate twenty randomly selected images from the SIR$^2$ dataset on a five-point scale regarding reflection removal and structure preservation. The mean opinion score (MOS) of the study is presented in Table \ref{tab:table}; our method has achieved the best MOS among all state-of-the-art methods. 
\begin{table}[]
\centering
\resizebox{0.45\textwidth}{!}{%
\begin{tabular}{|l|l|l|c|c|}
\hline
Method & \multicolumn{2}{c|}{Metric} & \multicolumn{2}{c|}{MOS} \\ \hline
 & PSNR & SSIM & \begin{tabular}[c]{@{}c@{}}Reflection \\ Removal\end{tabular} & \begin{tabular}[c]{@{}c@{}}Structure \\ Preservation\end{tabular} \\ \hline
PNet \cite{zhang2018single} & 20.33 & 0.833 & 3.9 & 3.5 \\ \hline
ERRNet \cite{wei2019single} & 23.66 & 0.870 & 3.6 & 3.8 \\ \hline
GCNet \cite{abiko2019single} & 22.82 & 0.926 & 3.2 & 3.3 \\ \hline
IBCLN \cite{li2020single} & 24.32 & 0.884 & 3.3 & 3.5 \\ \hline
CoRRN \cite{wan2019corrn} & 24.19 & 0.903 & 3.8 & 3.4 \\ \hline
Proposed & 24.52 & 0.927 & 4.2 & 4.0 \\ \hline
\end{tabular}%
}
\caption{Quantitative evaluation results using  PSNR,  SSIM and Mean Opinion Score (MOS). The result of objective evaluations are obtained by averaging the metric scores of all images from SIR$^2$ dataset.}
\label{tab:table}
\end{table}

Next, we present some of the reflection removal results for subjective evaluation in Figure \ref{fig:sota}. In the top row of  Figure \ref{fig:sota} an example image from the wild screen subset is shown. The corresponding estimated transmission image of the different methods is shown on the right side. Wild screen set contains outdoor reflection images with complicated environment. Here we can see reflection removal result of our method is favorably better than the other state of the art method. Noticeably, ERRNet, IBCLN, completely failed to remove the reflection effect. Next, we considered reflection images from the postcard subset. This subset contains many challenging examples with different settings in a controlled environment. In the first example, we took a reflection image taken by the camera with an aperture size of F11 and shutter speed 1/3s. In this example, we can see, CoRRN is the least performer in terms of reflection removal. Non arguably, our method has removed the reflection effect most efficiently. Next, we consider another image of the same object but taken by the camera with an aperture size of F32 and shutter speed 3s. Big aperture size makes the reflection more vital \cite{wan2017benchmarking}; as shown in 3rd-row images of Figure \ref{fig:sota}. This time ERRNet, GCNet, IBCLN again failed to remove reflection; however, they decently removed reflection in the previous setting with low reflection strength. The reflection removal result of CoRRN and PNet is lower than the proposed method for this example also. Our iterative method makes it possible to remove strong reflection from the background by iteratively making the background more accurate than the previously estimated one. To further concrete our claim, we again took another challenging example from the solid object subset. Here the reflection image is taken through a thick glass of 5mm. Thick glass creates a ghosting cue effect in the reflection image \cite{shih2015reflection}. We can see that most of the state of the art methods failed to remove the ghosting cue effect, whereas our method has acceptably removed reflection from the image. Next, we took another reflection image from the wild screen subset with a very strong reflection effect. Removing reflection from these mixture images is challenging due to the complex type of reflection; our method has still performed favorably better than the other methods. We found different methods perform satisfactorily in removing specific reflection types but  exhibited limitations when the kind of reflection changes. These methods either considered a distinct image formal model while designing the loss function, regularization constraint, and choosing the prior or are completely data-driven. Whereas, our data-driven method utilized special features of the gradient images, but without any implicit consideration of the reflection formation model, which makes our method better generalized over diverse reflection images. In the supplementary material, we have given more results of reflection removal of our method.

\section{Conclusion}
\label{sec:illust}

This paper presents a novel method for single image reflection removal. Specifically, we proposed an iterative gradient encoding network and a feature co-occurrence loss. Unlike the conventional pipeline, we did not use any gradient inference network but used an encoding network to use the features of the gradient for understanding the pixel location of the transmission layer. Comprehensive experiments on the benchmark dataset of SIR$^2$ validates that our method is more competent in removing strong reflection than the existing state of the art method. We have evaluated our method in reflection images taken in a controlled environment with different settings. Our method satisfactorily performed in all settings, unlike the previous method, which failed to perform in all settings with the same competency. 

\bibliographystyle{IEEEbib}
\bibliography{strings,refs}

\begin{thebibliography}{10}

\bibitem{guo2014robust}
Xiaojie Guo, Xiaochun Cao, and Yi~Ma,
\newblock ``Robust separation of reflection from multiple images,''
\newblock in {\em Proceedings of the IEEE Conference on Computer Vision and
  Pattern Recognition}, 2014, pp. 2187--2194.

\bibitem{sarel2004separating}
Bernard Sarel and Michal Irani,
\newblock ``Separating transparent layers through layer information exchange,''
\newblock in {\em European Conference on Computer Vision}. Springer, 2004, pp.
  328--341.

\bibitem{szeliski2000layer}
Richard Szeliski, Shai Avidan, and Padmanabhan Anandan,
\newblock ``Layer extraction from multiple images containing reflections and
  transparency,''
\newblock in {\em Proceedings IEEE Conference on Computer Vision and Pattern
  Recognition. CVPR 2000 (Cat. No. PR00662)}. IEEE, 2000, vol.~1, pp. 246--253.

\bibitem{li2014single}
Yu~Li and Michael~S Brown,
\newblock ``Single image layer separation using relative smoothness,''
\newblock in {\em Proceedings of the IEEE Conference on Computer Vision and
  Pattern Recognition}, 2014, pp. 2752--2759.

\bibitem{shih2015reflection}
YiChang Shih, Dilip Krishnan, Fredo Durand, and William~T Freeman,
\newblock ``Reflection removal using ghosting cues,''
\newblock in {\em Proceedings of the IEEE Conference on Computer Vision and
  Pattern Recognition}, 2015, pp. 3193--3201.

\bibitem{li2020single}
Chao Li, Yixiao Yang, Kun He, Stephen Lin, and John~E Hopcroft,
\newblock ``Single image reflection removal through cascaded refinement,''
\newblock in {\em Proceedings of the IEEE/CVF Conference on Computer Vision and
  Pattern Recognition}, 2020, pp. 3565--3574.

\bibitem{levin2007user}
Anat Levin and Yair Weiss,
\newblock ``User assisted separation of reflections from a single image using a
  sparsity prior,''
\newblock {\em IEEE Transactions on Pattern Analysis and Machine Intelligence},
  vol. 29, no. 9, pp. 1647--1654, 2007.

\bibitem{fan2017generic}
Qingnan Fan, Jiaolong Yang, Gang Hua, Baoquan Chen, and David Wipf,
\newblock ``A generic deep architecture for single image reflection removal and
  image smoothing,''
\newblock in {\em Proceedings of the IEEE International Conference on Computer
  Vision}, 2017, pp. 3238--3247.

\bibitem{wan2018crrn}
Renjie Wan, Boxin Shi, Ling-Yu Duan, Ah-Hwee Tan, and Alex~C Kot,
\newblock ``Crrn: Multi-scale guided concurrent reflection removal network,''
\newblock in {\em Proceedings of the IEEE Conference on Computer Vision and
  Pattern Recognition}, 2018, pp. 4777--4785.

\bibitem{lyu2019reflection}
Youwei Lyu, Zhaopeng Cui, Si~Li, Marc Pollefeys, and Boxin Shi,
\newblock ``Reflection separation using a pair of unpolarized and polarized
  images,''
\newblock in {\em Advances in neural information processing systems}, 2019, pp.
  14559--14569.

\bibitem{sun2019multi}
Jun Sun, Yakun Chang, Cheolkon Jung, and Jiawei Feng,
\newblock ``Multi-modal reflection removal using convolutional neural
  networks,''
\newblock {\em IEEE Signal Processing Letters}, vol. 26, no. 7, pp. 1011--1015,
  2019.

\bibitem{abiko2019single}
Ryo Abiko and Masaaki Ikehara,
\newblock ``Single image reflection removal based on gan with gradient
  constraint,''
\newblock in {\em Asian Conference on Pattern Recognition}. Springer, 2019, pp.
  609--624.

\bibitem{wan2019corrn}
Renjie Wan, Boxin Shi, Haoliang Li, Ling-Yu Duan, Ah-Hwee Tan, and Alex~Kot
  Chichung,
\newblock ``Corrn: Cooperative reflection removal network,''
\newblock {\em IEEE Transactions on Pattern Analysis and Machine Intelligence},
  2019.

\bibitem{gatys2015neural}
Leon~A Gatys, Alexander~S Ecker, and Matthias Bethge,
\newblock ``A neural algorithm of artistic style,''
\newblock {\em arXiv preprint arXiv:1508.06576}, 2015.

\bibitem{wan2017benchmarking}
Renjie Wan, Boxin Shi, Ling-Yu Duan, Ah-Hwee Tan, and Alex~C Kot,
\newblock ``Benchmarking single-image reflection removal algorithms,''
\newblock in {\em Proceedings of the IEEE International Conference on Computer
  Vision}, 2017, pp. 3922--3930.

\bibitem{wan2016depth}
Renjie Wan, Boxin Shi, Tan~Ah Hwee, and Alex~C Kot,
\newblock ``Depth of field guided reflection removal,''
\newblock in {\em 2016 IEEE International Conference on Image Processing
  (ICIP)}. IEEE, 2016, pp. 21--25.

\bibitem{simonyan2014very}
Karen Simonyan and Andrew Zisserman,
\newblock ``Very deep convolutional networks for large-scale image
  recognition,''
\newblock {\em arXiv preprint arXiv:1409.1556}, 2014.

\bibitem{zhang2018single}
Xuaner Zhang, Ren Ng, and Qifeng Chen,
\newblock ``Single image reflection removal with perceptual losses,''
\newblock CVPR, 2018.

\bibitem{wei2019single}
Kaixuan Wei, Jiaolong Yang, Ying Fu, David Wipf, and Hua Huang,
\newblock ``Single image reflection removal exploiting misaligned training data
  and network enhancements,''
\newblock in {\em Proceedings of the IEEE Conference on Computer Vision and
  Pattern Recognition}, 2019, pp. 8178--8187.

\end{thebibliography}

\end{document}